\documentclass[aps,prl,superscriptaddress,twocolumn,showpacs,preprintnumbers,amsmath,amssymb]{revtex4}
\usepackage[dvips]{graphicx,color}% Include figure files
\usepackage{dcolumn}% Align table columns on decimal point
\usepackage{bm}% bold math
\usepackage{ulem}
\usepackage{soul}
\begin{document}

\title{Strong Suppression of the Spin Hall Effect in the Spin Glass State}

\author{Y. Niimi}
\email{niimi@phys.sci.osaka-u.ac.jp}
\altaffiliation{Present address: Department of Physics, Osaka University, Toyonaka, Osaka 560-0043, Japan}
\affiliation{Institute for Solid State Physics, University of Tokyo, 5-1-5 Kashiwa-no-ha, Kashiwa, Chiba 277-8581, Japan}
%\affiliation{Department of Physics, Osaka University, 1-1 Machikaneyama, Toyonaka Osaka 560-0043, Japan}
\author{M. Kimata}
\affiliation{Institute for Solid State Physics, University of Tokyo, 5-1-5 Kashiwa-no-ha, Kashiwa, Chiba 277-8581, Japan}
\author{Y. Omori}
\affiliation{Institute for Solid State Physics, University of Tokyo, 5-1-5 Kashiwa-no-ha, Kashiwa, Chiba 277-8581, Japan}
\author{B. Gu}
\affiliation{Advanced Science Research Center, Japan Atomic Energy Agency, Tokai 319-1195, Japan}
\author{T. Ziman}
\affiliation{Institut Laue Langevin, Bo\^\i te Postale 156, F-38042 Grenoble Cedex 9, France}
\affiliation{LPMMC (UMR 5493), Universit\'{e} Grenoble Alpes and CNRS, 25 rue des Martyrs, B.P. 166, 38042 Grenoble, France}
\author{S. Maekawa}
\affiliation{Advanced Science Research Center, Japan Atomic Energy Agency, Tokai 319-1195, Japan}
\affiliation{ERATO, Japan Science and Technology Agency, Sendai 980-8577, Japan}
\author{A. Fert}
\affiliation{Unit\'{e} Mixte de Physique CNRS/Thales, 91767 Palaiseau France associ\'{e}e \`{a} l'Universit\'{e} de Paris-Sud, 91405 Orsay, France}
\author{Y. Otani}
\affiliation{Institute for Solid State Physics, University of Tokyo, 5-1-5 Kashiwa-no-ha, Kashiwa, Chiba 277-8581, Japan}
\affiliation{RIKEN-CEMS, 2-1 Hirosawa, Wako, Saitama 351-0198, Japan}

\date{\today}

\begin{abstract}
We have measured spin Hall effects in spin glass metals, CuMnBi alloys, with 
the spin absorption method in the lateral spin valve structure. 
Far above the spin glass temperature $T_{g}$ where 
the magnetic moments of Mn impurities are randomly frozen, 
the spin Hall angle of CuMnBi ternary alloy 
is as large as that of CuBi binary alloy. 
Surprisingly, however, it starts to decrease at about 4$T_{g}$ and 
becomes as little as 7 times smaller at $0.5 T_{g}$. 
A similar tendency was also observed in 
anomalous Hall effects in the ternary alloys. 
We propose an explanation in terms of a simple model 
considering the relative dynamics between 
the localized moment and the conduction electron spin.
\end{abstract}

% PACS, the Physics and Astronomy
                             % Classification Scheme.
%\keywords{Suggested keywords}%Use showkeys class option if keyword
                              %display desired

\maketitle

Spin glass is one of the magnetic ordering phases 
with very complex structures, and has been studied 
for several decades~\cite{rmp_1986}. 
It typically appears when magnetic impurities are 
randomly distributed in a nonmagnetic host metal. 
Below a certain temperature, 
so-called spin glass temperature $T_{g}$, 
magnetic moments at the impurity sites start to order, 
but since their spatial distribution is random, the 
Ruderman-Kittel-Kasuya-Yosida interactions 
between the spins mediated by conduction electrons are also random. 
Consequently, the ground state of the spin glass is not a simple phase such as 
ferromagnet or antiferromagnet 
but these two are intricately distributed and 
the randomness induces a large frustration between the spins. 
Since the spin glass can be regarded as a model system 
of information science and also brain~\cite{monasson_nature_1999}, 
it is important to understand the spin glass system more deeply. 

Among several spin glass materials, 
Mn-doped Cu (CuMn) is one of the typical spin glass systems and 
has been studied mainly by magnetization 
measurements~\cite{nagata_prb_1979,monod_jp_1980,morgownik_prb_1981}. 
The magnetic susceptibility shows a typical cusp at $T_{g}$ 
under zero field cooling (ZFC), and it is constant under field cooling (FC). 
However, some fundamental questions still remain unsolved.
While the mangetization is very sensitive to 
the applied magnetic filed, the transport properties are quite robust 
for the field~\cite{levy_prl_1991,capron_prl_2013}, which is different from 
another typical spin glass metal, AuFe, 
where the magnetization is proportional to the Hall 
resistivity~\cite{vloeberghs_epl_1990,petit_prl_2002,taniguchi_prl_2004}. 
%and a much larger Dzyaloshinsky-Moriya (DM) anisotropy field is 
%obtained compared to CuMn. 
Moreover, not only the complex ground state in the spin glass 
but also spin fluctuations related to 
the spin chirality~\cite{tatara_jpsj_2002} 
have not been fully understood yet. 
To reveal such properties in the spin glass systems, 
another type of measurement is highly desirable. 

In this Letter, we present spin transport measurements in CuMn. 
Among several types of spin transport measurements, 
here we chose spin Hall effect (SHE) measurements 
using the spin absorption method in the lateral spin valve structure. 
This method enables us to estimate quantitatively 
the spin Hall (SH) angle, 
conversion yield between charge and spin currents, 
and the spin diffusion length on the same 
device~\cite{niimi_prl_2011,niimi_prl_2012,niimi_prb_2014}. 
As a matter of fact, CuMn does not show a clear 
SHE signal because Mn is not a good scatterer 
for the spin current~\cite{supplemental_material,fert_jmmm_1981}. 
Thus, we added a heavy metal impurity 
in CuMn~\cite{fert_jmmm_1981,fert_prl_1980,levy_prb_1981,bouchiat_jjap_1987}. 
In the present work, we measured the SHE in CuMnBi ternary alloy. 
We have already shown that Bi in Cu works as a very good skew scatterer 
and the SH angle of CuBi is very large~\cite{niimi_prl_2012}. 
When a pure spin current, flow of only the spin angular momentum, 
is injected into the ternary alloy, 
it is converted into a charge current 
at the Bi impurity site through the inverse process of SHE. 
The converted charge current also feels spin fluctuations at the 
Mn sites [see Fig.~1(a)]. 
Surprisingly, the SH angle of CuMnBi starts to decrease 
at about 4$T_{g}$ and becomes 7 times smaller compared to that of CuBi 
at 0.5$T_{g}$. 
This reduction stems from randomized directions 
of conduction electron spins due to the fluctuating Mn moments. 

\begin{figure}
\begin{center}
\includegraphics[width=8.5cm]{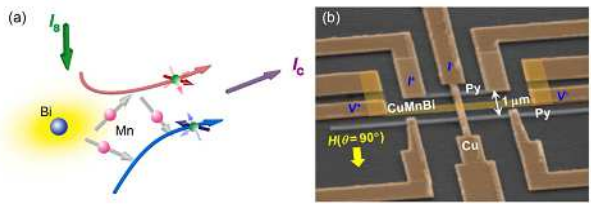}
\caption{(Color online) (a) Illustration of ISHE in CuMnBi ternary alloy. A pure spin current $I_{\rm S}$ is converted into a charge current $I_{\rm C}$ at the Bi site. Red and blue arrows with green spheres are spins of conduction electrons ($|e|$) and the shadows indicate that the conduction electron spins are randomized by the localized moments at the Mn sites. The curved arrows show the motions of spin-up and spin-down electrons. (b) Scanning electron micrograph of a typical device. The current leads and voltage probes are for the ISHE measurement.} \label{fig1}
\end{center}
\end{figure}

We prepared two types of devices to evaluate the SHE in the 
spin glass system. 
One is for the SHE measurement using the lateral spin valve and the 
other is for the anomalous Hall effect (AHE) measurement 
with a simple Hall bar structure. 
The latter measurement was originally performed by Fert $\textit{et al.}$ 
using CuMnX (X: transition metal) ternary alloys~\cite{fert_jmmm_1981}. 
In their case, a small amount of Mn was added in Cu. 
Thus, the interaction between the Mn impurities could be ignored, and 
the localized moments at the Mn sites simply followed a Curie law and 
worked as spin polarizers. 
In the present case, the concentration of Mn is 
much higher than in Ref.~\cite{fert_jmmm_1981}. 
As we will see later on, the Mn impurities work as spin polarizers 
above a certain temperature ($T^{*}$), while they interact with each other 
below this temperature. 

Figure~1(b) shows a scanning electron microscopy image of 
a typical SHE device. 
It consists of two ferromagnetic permalloy 
(Ni$_{81}$Fe$_{19}$, hereafter Py) wires and a CuMnBi middle wire. 
These three wires are bridged by a nonmagnetic Cu wire. 
Further details on the SHE device 
%sample fabrications and the dimensions of the wires 
should be referred to Supplemental Material~\cite{supplemental_material}.
In this work, we have fixed the concentration of Bi at 0.5\%, 
which shows the largest SHE signal among CuBi binary 
alloys~\cite{niimi_prl_2012}, 
and changed the concentration of Mn 
from 0 to 1.5\%. 
To check the reproducibility, six different devices have been 
measured for each Mn concentration.

We first mesured the inverse SHE (ISHE) and direct SHE (DSHE) in CuMnBi. 
When the electric current $I$ flows from the upper Py wire 
to the upper side of the Cu wire [see Fig.~1(b)], 
the resulting spin accumulation at the interface between the Py and Cu wires
induces a pure spin current (exactly same but opposite flows of 
spin-up and spin-down electrons) 
only on the lower side of the Cu wire. 
Most of the generated pure spin current is then absorbed vertically 
into the CuMnBi middle wire below Cu because of 
its stronger spin-orbit interaction. 
Both spin-up and spin-down electrons are deflected to 
the same direction by the ISHE, and a voltage 
is generated to prevent 
a charge current along the wire direction. 
By inverting the probe configuration (i.e., $I^{+} \Leftrightarrow V^{+}$, 
$I^{-} \Leftrightarrow V^{-}$), 
we can also measure the DSHE; 
with an electric current in the CuMnBi wire, the spin accumulation induced at 
the interface between Cu and CuMnBi can be detected as the nonlocal 
voltage between Py and Cu. For the DSHE measurement, 
the positive field is defined as the opposite direction to 
that in Fig.~1(b).

\begin{figure*}
\begin{center}
\includegraphics[width=11.5cm]{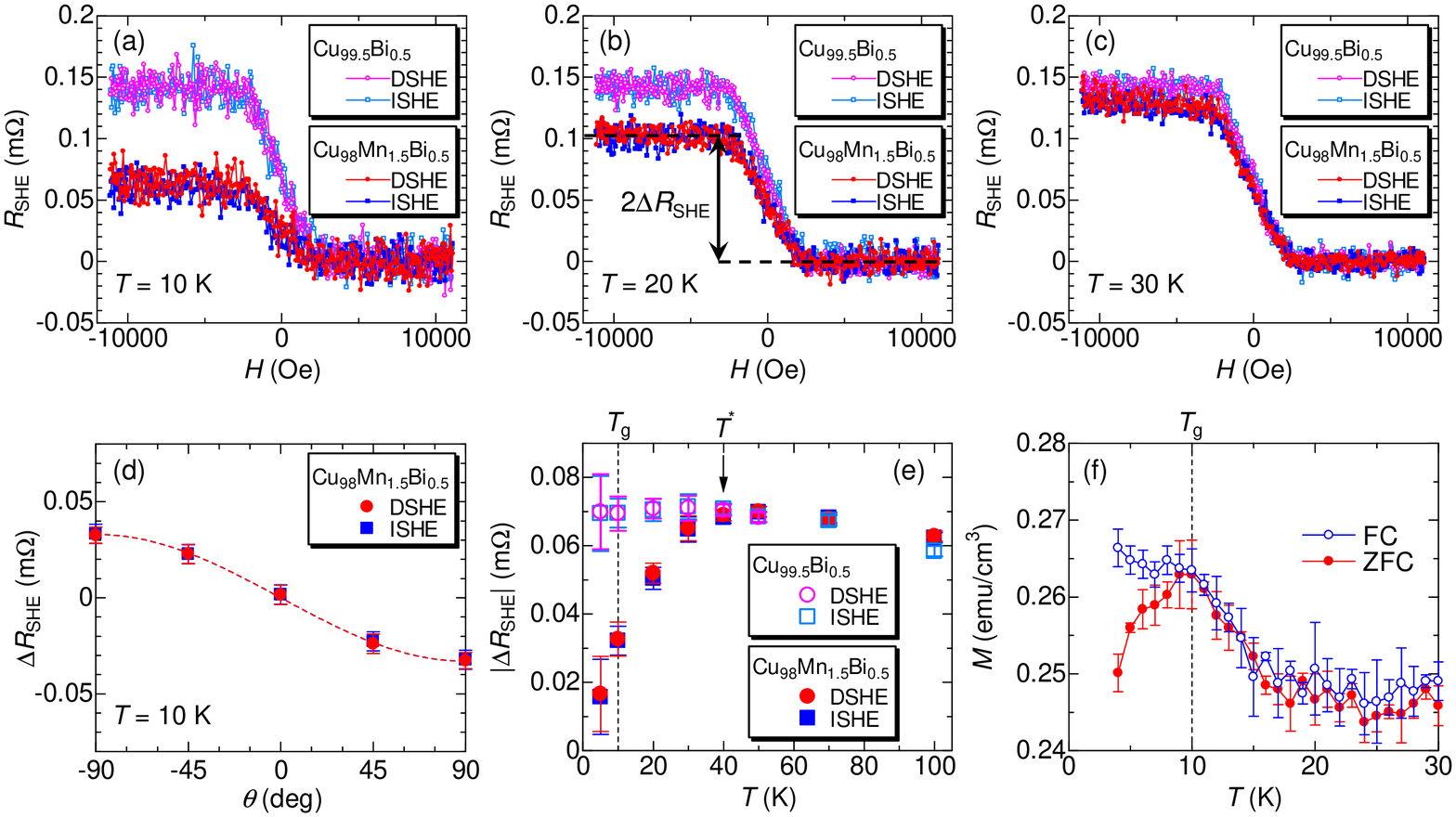}
\caption{(Color online) (a)-(c) ISHE (closed square) and DSHE (closed circle) resistances ($R_{\rm SHE}$) of Cu$_{98}$Mn$_{1.5}$Bi$_{0.5}$ measured at $T = 10$, 20, and 30~K. For comparision, $R_{\rm SHE}$ of Cu$_{99.5}$Bi$_{0.5}$ (open square and circle) are also plotted in the same figures. The amplitude of the SHE resistance $\Delta R_{\rm SHE}$ is defined in (b). Both $R_{\rm SHE}$ of Cu$_{98}$Mn$_{1.5}$Bi$_{0.5}$ and Cu$_{99.5}$Bi$_{0.5}$ are shifited along the vertical direction to see the difference of their amplitidues clearly. (d) Magnetic field angle ($\theta$) dependence of $\Delta R_{\rm SHE}$ of Cu$_{98}$Mn$_{1.5}$Bi$_{0.5}$ at $T = 10$~K. The broken curve shows $-|\Delta R_{\rm SHE}(\theta = 90^{\circ})| \sin \theta$. (e) $|\Delta R_{\rm SHE}|$ of Cu$_{98}$Mn$_{1.5}$Bi$_{0.5}$ (closed symbols) and Cu$_{99.5}$Bi$_{0.5}$ (open symbols) as a function of $T$. The vertical broken line indicates the spin glass temperature $T_{g}$ of Cu$_{98}$Mn$_{1.5}$Bi$_{0.5}$. The arrow shows the temperature ($T^{*}$) at which $|\Delta R_{\rm SHE}|$ starts to decrease. (f) Magnetizations of Cu$_{98}$Mn$_{1.5}$Bi$_{0.5}$ measured under ZFC (closed symbol) and FC (open symbol) at a small magnetic field ($H=50$~Oe) as a function of $T$. From the cusp position (the vertical broken line), $T_{g}$ can be determined.} \label{fig2}
\end{center}
\end{figure*}

In Figs.~2(a)-2(c), we show the ISHE and DSHE resistances 
($R_{\rm SHE} \equiv V/I$) 
of Cu$_{98}$Mn$_{1.5}$Bi$_{0.5}$ measured at 
$T = 10$, 20, and 30~K. 
$R_{\rm SHE}$ linearly changes with increasing the magnetic field $H$
and is saturated above 2000~Oe, which is the saturation field of 
the magnetization of 
the Py wire~\cite{niimi_prl_2011,niimi_prl_2012,niimi_prb_2014}.
At any temperature, both the ISHE and DSHE resistances have the same 
amplitude $\Delta R_{\rm SHE}$, which demonstrates the Onsager reciprocal 
relation in this system. 
We have also checked the field angle dependence of $\Delta R_{\rm SHE}$ 
in Fig.~2(d), and found that 
it follows a sinusoidal curve, 
typical of SHEs in nonmagnetic metals~\cite{niimi_prb_2014}. 
Most remarkable is the temperature dependence of 
$|\Delta R_{\rm SHE}|$. 
$|\Delta R_{\rm SHE}|$ of Cu$_{98}$Mn$_{1.5}$Bi$_{0.5}$ increases 
with increasing $T$, while that of Cu$_{99.5}$Bi$_{0.5}$
is basically constant up to 30~K. 
As summarized in Fig.~2(e), the former reaches the latter at 50~K, 
and both have the same amplitude above 50~K. 
This fact clearly shows that the large reduction in $|\Delta R_{\rm SHE}|$ 
of Cu$_{98}$Mn$_{1.5}$Bi$_{0.5}$ below $T^{*}=40$~K originates from 
the additional Mn impurities.

We next changed the Mn concentration down to 0\%. 
Figure~3(a) shows SHE resistivities $|\rho_{\rm SHE}^{\rm 3D}|$ 
of CuMnBi ternary alloys, 
obtained with a three-dimensional (3D) 
spin diffusion model~\cite{niimi_prl_2012,niimi_prb_2014}, 
divided by the resistivity induced by the Bi impurities $\rho_{\rm Bi}$. 
As demonstrated in our previous 
works~\cite{niimi_prl_2011,niimi_prl_2012,niimi_prb_2014},
in Cu-based alloys, 
$\rho_{\rm SHE}^{\rm 3D}/\rho_{\rm Bi}$ corresponds to 
the SH angle $\alpha_{\rm H}^{\rm 3D}$. 
Basically, $|\alpha_{\rm H}^{\rm 3D}|$ has the same tendency as 
$|\Delta R_{\rm SHE}|$: 
$|\alpha_{\rm H}^{\rm 3D}|$ of Cu$_{98}$Mn$_{1.5}$Bi$_{0.5}$ 
starts to decrease at $T^{*} = 40$~K, while 
$|\alpha_{\rm H}^{\rm 3D}|$ of Cu$_{99.5}$Bi$_{0.5}$ is constant. 
Interestingly, $|\alpha_{\rm H}^{\rm 3D}|$ of 
Cu$_{98}$Mn$_{1.5}$Bi$_{0.5}$ at 5~K is reduced by a factor of 7, 
compared to that at 50~K and also $|\alpha_{\rm H}^{\rm 3D}|$ of 
Cu$_{99.5}$Bi$_{0.5}$ at 5~K.
With decreasing the Mn concentration from 1.5 to 0\%, 
$T^{*}$ is shifted to the lower temperature side, 
and thus the total reduction of 
$|\alpha_{\rm H}^{\rm 3D}|$ at 5~K gets smaller. 

In order to relate the reduction of 
$|\alpha_{\rm H}^{\rm 3D}|$ with spin glass, 
we measured magnetizations $M$ of Cu$_{98}$Mn$_{1.5}$Bi$_{0.5}$ 
films under ZFC and FC in Fig.~2(f). 
A clear cusp was observed at $T = 10$~K in the ZFC measuement, while 
the magnetization was saturated for FC.
From the cusp position, we can determine $T_{g}$ 
of Cu$_{98}$Mn$_{1.5}$Bi$_{0.5}$ to be 10~K, 
which is consistent with that of 
CuMn binary alloys~\cite{nagata_prb_1979}. 
By combining the SHE and magnetization measurements of CuMnBi, 
we argue that the reduction of $|\alpha_{\rm H}^{\rm 3D}|$ already starts 
at 4 times higher temperature than $T_{g}$ 
(i.e., $T^{*} = 4T_{g}$) and still continues at $0.5T_{g}$. 

What is the origin of the large reduction of 
$|\alpha_{\rm H}^{\rm 3D}|$ below $T^{*}$? 
One possibility is related to the spin diffusion of 
conduction electrons in the spin glass state, 
as observed in electron spin resonance measurements 
with AgMn and 
CuMn~\cite{wu_prb_1985,mahdjour_zpb_1986,chien_jap_1994,j_phys_1989} 
where the spin relaxation of Mn moments is detected. 
However, this possibility can be ruled out for the following reasons: 
Based on the simple spin transport model 
in the skew scattering regime~\cite{takahashi_review_2008}, 
$\alpha_{\rm H} \propto 1/(\rho_{\rm M}\lambda_{\rm M})$ but 
$\lambda_{\rm M} \propto 1/\rho_{\rm M}$ where $\rho_{\rm M}$ and 
$\lambda_{\rm M}$ are the resistivity and 
the spin diffusion length of CuMnBi, respectively. 
Thus, $\alpha_{\rm H}$ is independent of those parameters. 
As can be seen in Fig.~3(b), 
the spin diffusion length $\lambda_{\rm M}^{\rm 3D}$ estimated from 
nonlocal spin valve measurements~\cite{niimi_prb_2014} 
decreases by a factor of 2 for Cu$_{98}$Mn$_{1.5}$Bi$_{0.5}$ 
as $T$ approaches $T_{g}$, and shows much less temperature dependence 
for other Mn concentrations. On the other hand, 
$\rho_{\rm M}$ shows a very small reduction (less than 1\%) below $T^{*}$ 
(see Fig.~S3 in Ref.~\cite{supplemental_material}). 
These temperature dependencies of 
$\lambda_{\rm M}$ and $\rho_{\rm M}$ 
completely fail to explain the large suppression 
of $\alpha_{\rm H}$ of CuMnBi below $T^{*}$.

The mechanism we then propose is that the relative dynamics 
of the polarization of the electron spin $\vec{s}$ and 
the localized moments leads to a random precession of $\vec{s}$.
This reduces the converted charge current $\vec{I}_{\rm C}$ 
($\propto  \vec{I}_{\rm S} \times \vec{s}$) 
strongly because of the vector product. 
At high temperatures, the Mn moments fluctuate quickly and 
for the spin current the precession is negligible. 
As the spin glass freezes, the dynamics slows and correlations of 
the Mn moments decay with a characteristic frequency $\nu(T)$ 
which vanishes as a power law as we approach $T_g$ from above, 
just as has been seen in experiments 
with neutrons and muons on bulk spin glasses~\cite{uemura_prb_1985}. 
This can be understood as the effect of motional narrowing. 

We now use a simple phenomenological model, as previously used 
to model the broadening of the conduction electron spin resonance 
in metallic spin glasses close to freezing~\cite{hou_prb_1984}. 
The polarization $\vec{s}$ precesses in an effective time-varying 
magnetic field $\vec{S}_{eff}(t)$, as well as a constant $\vec{S}_0$ 
that includes any applied external field: 
${\frac{\partial{\vec s}}{\partial t}}={\vec s}\times (\vec{S}_0+\vec{S}_{eff}(t))$. 
The instantaneous $\vec{S}_{eff}(t)$ is random 
in both direction and magnitude, 
with a distribution width proportional to the $s$-$d$ interaction, 
and fluctuates on a time scale of $\nu^{-1}(T)$. 
For an appropriate choice of $\vec{S}_{eff}(t)$, 
the integration of the time-dependent equation 
for $\vec{s}$ defines the Kubo-Toyabe model~\cite{hayano_prb_1979} 
as used in muon experiments. 
For all frequencies, the skew scattered current is reduced 
from the temperature independent $\alpha_{\rm H}$ to 
$\alpha_{\rm H}\left< s \right>$ 
where $\left< s \right> = G_{z}(t=\tau_{\rm sk},\nu=\nu(T))$. 
$G_{z}$ is the average spin correlation with respect to 
its initial polarization. 
During the skew scattering, that takes place in a time 
$t = \tau_{\rm sk}$~\cite{skew_time}, 
and the polarization precesses in $\vec{S}_{eff}$ 
that varies with the frequency 
$\nu(T)$. Well above $T_g$, we are in the motionally narrowed limit 
of large $\nu(T)$ where $G_{z}$ is 1. 
For lower temperatures, $G_{z}$ decreases with $\nu(T)$ and 
$\alpha_{\rm H}$ is reduced 
(see Fig.~S8 in Ref.~\cite{supplemental_material}).

As shown above, the SHE in spin glass depends strongly on $T$. 
However, it is quite robust for the applied magnetic field. 
In the SHE measurements, 
there is no difference between ZFC and FC even under 
$H = 1$~T. This fact looks inconsistent with 
magnetization measurements~\cite{nagata_prb_1979} 
but are consistent with the previous transport 
measurements~\cite{levy_prl_1991,capron_prl_2013}. In addition, 
recent SHE measurements \cite{wei_ncomm_2012,du_prb_2014} 
reveal that \textit{homogenous} 
magnetizations such as ferromagnetic and antiferromagnetic 
states are irrelevant for the amplitudes of SHE signals. 
This indicates that the \textit{spin fluctuations} severely affect 
$\alpha_{\rm H}(T)$ and the energy scale of the fluctuations 
is significantly larger than the applied field. 
The spin diffusion length, on the other hand, 
is also affected by the fluctuating fields, 
but is less sensitive to them than 
$\alpha_{\rm H}(T)$ [see Fig.~3(b)].

\begin{figure}
\begin{center}
\includegraphics[width=8.5cm]{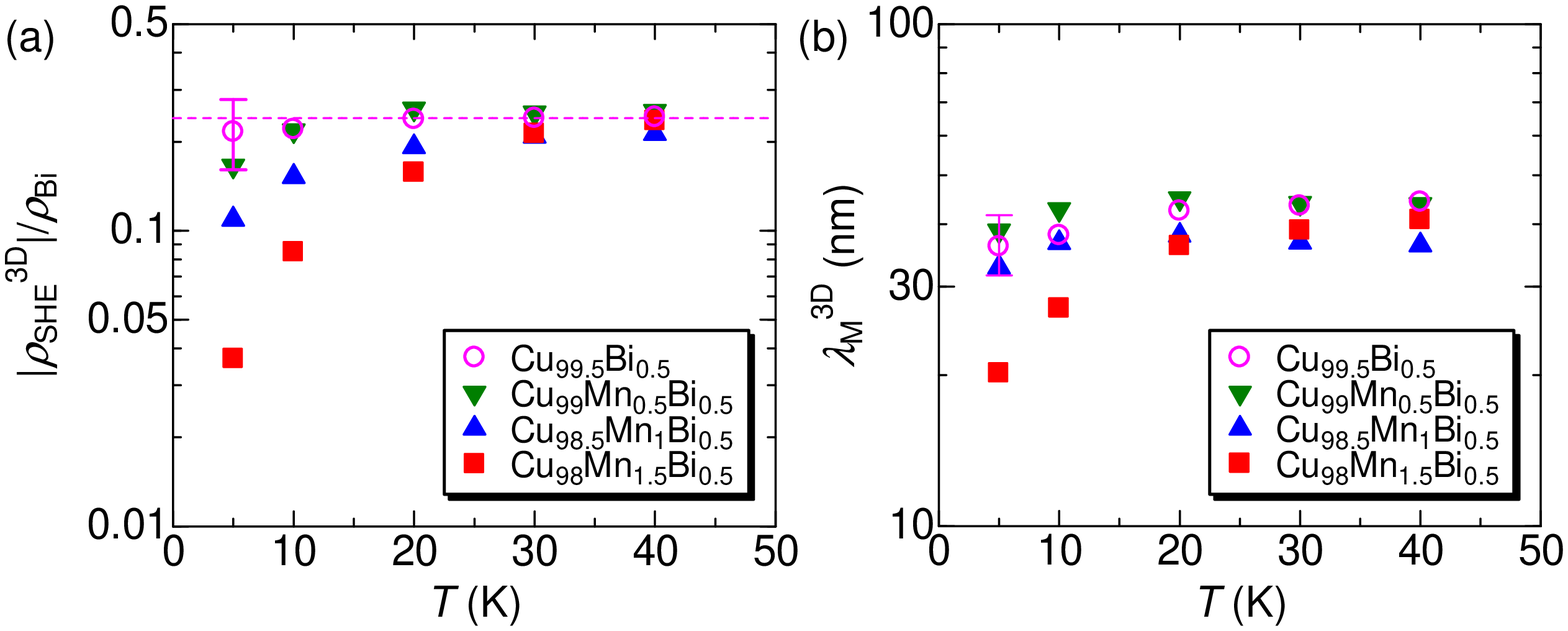}
\caption{(Color online) (a) SHE resistivities of CuMnBi $|\rho_{\rm SHE}^{\rm 3D}|$, obtained with the 3D calculation, divided by the resistivity induced by the Bi impurities $\rho_{\rm Bi}$ as a function of $T$. The Bi concentration is fixed at 0.5\%, while the Mn concentration is changed from 0 to 1.5\%. The broken line in the figure shows $|\alpha_{\rm H}^{\rm 3D}|$ of Cu$_{99.5}$Bi$_{0.5}$. (b) Spin diffusion lengths of CuMnBi $\lambda_{\rm M}^{\rm 3D}$ obtained with the 3D calculation as a function of $T$.} \label{fig3}
\end{center}
\end{figure}

To support our findings in the SHEs in CuMnBi, 
we have also performed the AHE measurements. 
When the Mn concentration is low enough, Mn works as a 
spin polarizer and its magnetization follows a simple 
Curie law, i.e., $M \propto T^{-1}$. 
On the other hand, Mn does not work as a 
skew scatterer~\cite{fert_jmmm_1981,supplemental_material}. 
Thus, to see the spin-dependent transport in Cu-based alloys, 
an additional metal with stronger spin-orbit interaction is needed, 
as detailed in Ref.~\cite{fert_jmmm_1981}.
Figure~4(a) shows a typical Hall resistivity of 
Cu$_{98}$Mn$_{1.5}$Bi$_{0.5}$ at $T = 50$~K. As reference signals, 
we also plot the Hall resistivities $\rho_{yx}$ of 
Cu$_{99.5}$Bi$_{0.5}$ and Cu$_{97}$Mn$_{3}$ at the same temperature. 
Only for Cu$_{98}$Mn$_{1.5}$Bi$_{0.5}$, an anomaly can be seen 
near 0~T.

The anomalous part $d\delta \rho_{yx}/dH$ can be extracted by subtracting the 
derivation of $\rho_{yx}$ at zero field from the 
one of normal Hall resistivity at 1~T. 
As demonstrated in Ref.~\cite{fert_jmmm_1981}, 
by plotting $d\delta \rho_{yx}/dH$ as a function of $1/T$, 
$\alpha_{\rm H}$ of CuBi can be evaluated (see the inset of Fig.~4(b) 
and Ref.~\cite{supplemental_material}). 
It is $-0.23(\pm 0.06)$, which is quantitatively consistent with 
$\alpha_{\rm H}^{\rm 3D}$ determined by the SHE device~\cite{niimi_prl_2012}. 
As we decrease $T$, the amplitude of $|d\delta \rho_{yx}/dH|$ 
increases inversely proportional to $T$, but 
it starts to decrease at the exactly same temperature 
as $T^{*}$ [see Fig.~4(b)]. 
A similar tendency can also be seen in CuMnIr ternary alloys 
(see Ref.~\cite{supplemental_material} for more details). 

Finally, let us discuss the difference between the two typical 
spin glass materials, CuMn and AuFe.
As mentioned in the introduction, 
in AuFe, both the magnetization and the AHE show 
the same temperature and field dependencies~\cite{taniguchi_prl_2004}, 
while such behavior cannot be seen in CuMn. 
This can be explained as follows. 
Mn has a magnetic moment but it does not function as a skew scatterer 
for conduction electron spins. 
Thus, an additional skew-scatterer ``X" is added and 
the interaction of the Mn moments on spin currents occurs via the X site. 
Such an indirect interaction makes the temperature $T^{*}$ where 
the SHE feels the effects of spin correlations.
In AuFe, on the other hand, the Fe impurity has both properties. 
With such an \textit{on-site} interaction,
$T^{*} = T_{g}$ and 
a clear difference between ZFC and FC can be seen 
even in the AHE restivities. 
Further investigations using different
combinations of host and impurity metals are needed to
unveil all the details.

\begin{figure}
\begin{center}
\includegraphics[width=8.5cm]{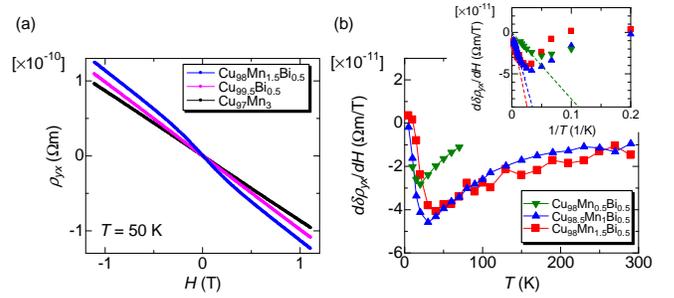}
\caption{(Color online) (a) Hall resistivities $\rho_{yx}$ of Cu$_{98}$Mn$_{1.5}$Bi$_{0.5}$, Cu$_{99.5}$Bi$_{0.5}$, and Cu$_{97}$Mn$_{3}$ measured with Hall bars at $T = 50$~K. %An anomaly can be seen near zero field only for Cu$_{98}$Mn$_{1.5}$Bi$_{0.5}$, while Cu$_{99.5}$Bi$_{0.5}$ and Cu$_{97}$Mn$_{3}$ show normal Hall resistances. 
(b) Differential values of anomalous parts $d\delta \rho_{yx}/dH$ for CuMnBi as a function of $T$. In the inset, they are plotted as a function of $1/T$. The broken lines show the linear fits to obtain $\alpha_{\rm H}$.} \label{fig4}
\end{center}
\end{figure}

In summary, we have studied the SHEs and AHEs 
in the spin glass systems using the CuMnBi ternary alloys. 
The SH angle $|\alpha_{\rm H}^{\rm 3D}|$ of Cu$_{98}$Mn$_{1.5}$Bi$_{0.5}$ 
at $T = 50$~K coincides with that of Cu$_{99.5}$Bi$_{0.5}$. 
With decreasing temperature, however, it starts to decrease 
at $T^{*} = 4T_{g}$ and becomes 7 times smaller at 0.5$T_{g}$. 
With decreasing concentrations of the Mn impurities, 
$T^{*}$ shifts to lower temperatures. 
These results suggest that the SHE 
could be exploited to probe fluctuating spin states in 
complex spin structures such as spin liquids.

We acknowledge fruitful discussions with K. Kobayashi, 
M. Ferrier, K. Tanabe, T. Arakawa, Y. Matsumoto, T. Kato, and H. Akai. 
We thank H. Mori, A. Ueda, J. Yoshida, and H. Idzuchi 
for their technical support on 
the magnetization measumrents, and Y. Iye and S. Katsumoto 
for the use of the lithography facilities. 
This work was supported by Grants-in-Aid for Scientific Research 
(No.24740217, No.23244071, and No. 60245610) and 
also by Foundation of Advanced Technology Institute.
%(Grant No. 22840012, 24740217, and 23244071).


\begin{thebibliography}{00}
\bibitem{rmp_1986}
K. Binder and A. P. Young, Rev. Mod. Phys. {\bf 58}, 801 (1986).
\bibitem{monasson_nature_1999}
R. Monasson, R. Zecchina, S. Kirkpatrick, B. Selman, and L. Troyansky, Nature (London) {\bf 400}, 133 (1999).
\bibitem{nagata_prb_1979}
S. Nagata, P. H. Keesom, and H. R. Harrison, Phys. Rev. B {\bf 19}, 1633 (1979).
\bibitem{monod_jp_1980}
J. J. Prejean, M. J. Joliclerc, and P. Monod, J. Physique {\bf 41}, 427 (1980).
\bibitem{morgownik_prb_1981}
A. F. J. Morgownik and J. A. Mydosh, Phys. Rev. B {\bf 24}, 5277 (1981).
\bibitem{levy_prl_1991}
P. G. N. de Vegvar, L. P. L\'evy, and T. A. Fulton, Phys. Rev. Lett. {\bf 66}, 2380 (1991)
\bibitem{capron_prl_2013}
T. Capron, G. Forestier, A. Perrat-Mabilon, C. Peaucelle, T. Meunier, C. B\"auerle, L. P. L\'evy, D. Carpentier, and L. Saminadayar, Phys. Rev. Lett. {\bf 111}, 187203 (2013).
\bibitem{vloeberghs_epl_1990}
H. Vloeberghs, J. Vranken, C. Van Haesendonck, and Y. Bruynseraede, Europhys. Lett. {\bf 12}, 557 (1990).
\bibitem{petit_prl_2002}
D. Petit, L. Fruchter, and I. A. Campbell, Phys. Rev. Lett. {\bf 88}, 207206 (2002).
\bibitem{taniguchi_prl_2004}
T. Taniguchi, K. Yamanaka, H. Sumioka, T. Yamazaki, Y. Tabata, and S. Kawarazaki, Phys. Rev. Lett. {\bf 93}, 246605 (2004).
\bibitem{tatara_jpsj_2002}
G. Tatara and H. Kawamura, J. Phys. Soc. Jpn. {\bf 71}, 2613 (2002).
\bibitem{niimi_prl_2011}
Y. Niimi, M. Morota, D. H. Wei, C. Deranlot, M. Basletic, A. Hamzic, A. Fert, and Y. Otani, Phys. Rev. Lett. {\bf 106}, 126601 (2011).
\bibitem{niimi_prl_2012}
Y. Niimi, Y. Kawanishi, D. H. Wei, C. Deranlot, H. X. Yang, M. Chshiev, T. Valet, A. Fert, and Y. Otani, Phys. Rev. Lett. {\bf 109}, 156602 (2012).
\bibitem{niimi_prb_2014}
Y. Niimi, H. Suzuki, Y. Kawanishi, Y. Omori, T. Valet, A. Fert, and Y. Otani, Phys. Rev. B {\bf 89}, 054401 (2014). 
\bibitem{supplemental_material}
See Supplemental Material at http://link.aps.org/
supplemental/10.1103/PhysRevLett.115.196602 
for sample preparation and for extra data on the SHE in CuMnBi,
which includes Refs.~\cite{wakamura_prl_2014}-\cite{fert_prl_2011}. 
\bibitem{wakamura_prl_2014} 
T. Wakamura, N. Hasegawa, K. Ohnishi, Y. Niimi, and Y. Otani, Phys. Rev. Lett. {\bf 112}, 036602 (2014).
\bibitem{buttiker_prb_2012}
P. Jacquod, R. S. Whitney, J. Meair, and M. B\"uttiker, Phys. Rev. B {\bf 86}, 155118 (2012).
\bibitem{fert_jphys_1973}
A. Fert, J. Phys. F: Met. Phys. {\bf 3}, 2126 (1973).
\bibitem{fert_prl_2011}
A. Fert and P. M. Levy, Phys. Rev. Lett. {\bf 106}, 157208 (2011).
\bibitem{fert_jmmm_1981}
A. Fert, A. Friederich, and A. Hamzic, J. Magn. Magn. Mat. {\bf 24}, 231 (1981).
\bibitem{fert_prl_1980}
A. Fert and P. M. Levy, Phys. Rev. Lett. {\bf 44}, 1538 (1980).
\bibitem{levy_prb_1981}
P. M. Levy and A. Fert, Phys. Rev. B {\bf 23}, 4667 (1981).
\bibitem{bouchiat_jjap_1987}
H. Bouchiat, N. de Courtenay, P. Monod, M. Ocio, and P. Refregier, Jpn. J. Appl. Phys. {\bf 26} Suppl. 26-3, 1951 (1987).
\bibitem{wu_prb_1985}
W.-Y. Wu, G. Mozurkewich, and R. Orbach, Phys. Rev. B {\bf 31}, 4557 (1985).
\bibitem{mahdjour_zpb_1986}
H. Mahdjour, C. Pappa, R. Wendler, and K. Baberschke, Z. Phys. B:Condens. Matter {\bf 63}, 351 (1986).
\bibitem{chien_jap_1994}
D. L. Leslie-Pelecky, F. VanWijland, C. N. Hoff, J. A. Cowen, A. Gavrin, and C.-L. Chien, J. Appl. Phys. {\bf 75}, 6489 (1994). 
\bibitem{j_phys_1989}
A. H. El-Sayed, S. Hedewy, and A. El-Samahy, J. Phys.:Condens. Matter {\bf 1}, 10515 (1989).
%\bibitem{tanaka_prb_2008}
%T. Tanaka, H. Kontani, M. Naito, T. Naito, D. S. Hirashima, K. Yamada, and J. Inoue, Phys. Rev. B {\bf 77}, 165117 (2008).
%\bibitem{hoffmann_prl_2010}
%O. Mosendz, J. E. Pearson, F. Y. Fradin, G. E. W. Bauer, S. D. Bader, and A. Hoffmann, Phys. Rev. Lett. {\bf 104}, 046601 (2010).
%\bibitem{morota_prb_2011}
%M. Morota, Y. Niimi, K. Ohnishi, D. H. Wei, T. Tanaka, H. Kontani, T. Kimura, and Y. Otani, Phys. Rev. B {\bf 83}, 174405 (2011).
%\bibitem{gu_prb_2012}
%B. Gu, T. Ziman, and S. Maekawa, Phys. Rev. B {\bf 86}, 241303(R) (2012).
%\bibitem{fujiki_ptp_1981}
%S. Fujiki and S. Katsura, Prog. Theor. Phys. {\bf 65}, 1130 (1981).
%\bibitem{fert_prl_2011}
%A. Fert and P. M. Levy, Phys. Rev. Lett. {\bf 106}, 157208 (2011).
\bibitem{takahashi_review_2008}
S. Takahashi and S. Maekawa, Sci. Tech. Adv. Mater. {\bf 9}, 014105 (2008).
\bibitem{uemura_prb_1985}
Y. J. Uemura, T. Yamazaki, D. R. Harshman, M. Senba, E. J. Ansaldo, Phys. Rev. B {\bf 31}  546 (1985).
\bibitem{hou_prb_1984}
M.-K. Hou, M. B. Salamon, and T. A. L. Ziman, Phys. Rev. B {\bf 30}, 5239 (1984).
\bibitem{hayano_prb_1979}
R. S. Hayano, Y. J. Uemura, J. Imazato, N. Nishida, T. Yamazaki, and R. Kubo, Phys. Rev. B {\bf 20}, 850 (1979).
\bibitem{skew_time}
$\tau_{\rm sk}$ is defined as $m/(ne^{2}\rho_{\rm SHE}^{\rm 3D})$ where $m$ and $n$ are the mass of the conduction electron and the electron density, respectively.
\bibitem{wei_ncomm_2012}
D. H. Wei, Y. Niimi, B. Gu, T. Ziman, S. Maekawa, and Y. Otani, Nat. Commn. {\bf 3}, 1058 (2012).
\bibitem{du_prb_2014}
C. Du, H. Wang, F. Yang, and P. C. Hammel, Phys. Rev. B {\bf 90}, 140407(R) (2014).
\end{thebibliography}
\end{document}